# *Field-induced spin-flip and spin-flop transitions in NdFeO₃*

M. M. Gomes[1], R. Vilarinho[1], E. Miranda[1], A. S. Silva[1], C. Kadlec[2], F. Kadlec[2], M. Lebeda[2], P. Proschek[3], M. Mihalik jr.[4], M. Mihalik[4], D. Jana[5], F. Choueikani[6], C. Faugeras[5], J. A. Paixão[7], E. de Prado[2], S. Kamba[2] and J. Agostinho Moreira[1,*]

1 – IFIMUP, Departamento de Física e Astronomia da Faculdade de Ciências, Universidade do Porto, Rua do Campo Alegre s/n, 4169-007 Porto, Portugal.

2 – Institute of Physics of the Czech Academy of Sciences, Na Slovance 2, 182 00 Prague 8, Czech Republic.

3 – Faculty of Mathematics and Physics, Charles University, Ke Karlovu 5, 121 Prague 2. Czech Republic.

4 – Institute of Experimental Physics Slovak Academy of Sciences, Watsonova 47, Košice, Slovak Republic.

5 – Laboratoire National des Champs Magnétiques Intenses, LNCMI-EMFL, CNRS UPR3228, Université Grenoble Alpes, Université Toulouse, Université Toulouse 3, INSA-T, Grenoble and Toulouse, France.

6 – Synchrotron SOLEIL, L'Orme des Merisiers, Saint Aubin – BP48, 91192 Gif-sur-Yvette, France.

7 – CFisUC, Department of Physics, University of Coimbra, P-3004-516 Coimbra, Portugal.

*Corresponding author: jamoreir@fc.up.pt

**Abstract**

Magnetic control of correlated spin systems is central to the development of next-generation spin-based technologies. Rare-earth orthoferrites provide an interesting platform in which exchange coupling between rare-earth 4f and transition-metal 3d moments generates competing magnetic interactions and multiple metastable states. Here, we show that the orientation of the applied magnetic field drives different magnetic phase transition sequences in NdFeO₃ across a broad temperature range. Using Raman and polarized terahertz spectroscopies, supported by magnetization and specific-heat measurements, we track the temperature- and field-dependent evolution of the different magnetic phases and the successive spin rearrangements, driven by 4f–3d magnetic anisotropic interactions. For fields applied along the crystallographic *c*-axis, a spin-reorientation transition is followed by spin-flop and spin-flip processes, producing an unexpectedly complex magnetic phase sequence at low temperatures. Below 8 K, precursor effects associated with ordering of the Nd-sublattice strongly modify the transition pathway. Our results demonstrate how anisotropic 4f–3d coupling enables magnetic-field control of coupled spin excitations and provide a route to accessing novel spin configurations in rare-earth orthoferrites.

## 1. Introduction

Materials with multiple interacting magnetic sublattices can exhibit a wide variety of spin configurations, providing a platform for both fundamental studies and applications in spintronics, magnonics, and high-frequency devices [1,2]. Rare-earth orthoferrites ($R$FeO$_3$) are prototypical systems in which interactions between Fe$^{3+}$-3$d$ and $R^{3+}$-4$f$ magnetic moments underlie complex, metastable magnetic states highly sensitive to external stimuli. These compounds crystallize in the orthorhombically distorted perovskite structure with space group $P$bnm, and feature two distinct magnetic sublattices associated with the $R^{3+}$ and Fe$^{3+}$ cations [2,3]. The interplay between these sublattices, combined with the magnetic anisotropic character of the $R^{3+}$ 4$f$-orbitals, lead to their rich magnetic behaviour, including spin-reorientation transitions, magnetization compensation, and metastable states [2–4]. Among orthoferrites, NdFeO$_3$ exhibits all of these phenomena over a broad temperature range, and serves as a prototypical system for studying coupled 4$f$–3$d$ spin dynamics [5,6].

NdFeO$_3$ undergoes a paramagnetic-to-antiferromagnetic phase transition at $T_{N1}$ = 760 K, associated with Fe magnetic sublattice ordering [7]. Due to the antisymmetric Dzyaloshinskii–Moriya interaction, Fe$^{3+}$ spins adopt a slightly canted configuration (canting angle is 8.5 mrad [8]), producing a weak ferromagnetic vector $F$ along the $c$-axis and an antiferromagnetic vector $G$ along the $a$-axis (G$_x$F$_z$ in Bertaut notation; $x \parallel a$, $y \parallel b$, $z \parallel c$), described by the Γ$_4$ irreducible representation of the $m'm'm$ magnetic point group [9,10]. This yields the magnetic space group $Pb'n'm$, stable down to $T_1$ = 170 K [4,11,12]. Below $T_1$, a spin-reorientation transition (SRT) occurs, with $F$ continuously rotating within the $ac$-plane, corresponding to the Γ$_{24}$ irreducible representation [9,10,13,14] and the predicted magnetic space group $P2_1'/n'$ [11,12]. The rotation completes at $T_2$ = 105 K, with $F$ aligned along $a$-axis (F$_x$G$_z$), and the low-temperature spin structure, down to $T_{N2}$ = 1.05 K, is described by Γ$_2$ representation of the $mm'm'$ point group, with space group $Pbn'm'$ [4,11,15].

Above $T_{N2}$ = 1.05 K, the Nd-sublattice aligns paramagnetically but polarized antiparallelly to the Fe-sublattice weak ferromagnetic component [16–18]. This is due to the dominant Fe-Nd exchange interaction and a weak Nd-Nd exchange term [16–18]. On cooling, the Fe-Nd strength increases and drives the spin reorientation transition [16,17]. The net magnetization along the $a$-axis peaks at $T_2$, then decreases and vanishes at $T_{comp}$ = 7.6 K, reversing direction upon further cooling [5]. This behaviour has been attributed to the increase in the Nd-sublattice magnetization as Nd-Fe exchange field strengthens [5]; however, direct experimental confirmation is still missing. Cooperative ordering of the $Nd^{3+}$ moments occurs only below $T_{N2}$, ascertained by the sharp peak in the specific heat as a function of temperature [19,20]. Yet, neutron diffraction experiments evidence the onset of magnetic diffraction peaks below 25 K, indexed to the $Nd^{3+}$ magnetic sublattice, and attributed to the polarization of the electronic orbital magnetic momenta [20]. However, a C-type ordering of the Nd-sublattice and a deviation from the pure G-type ordering of the Fe-sublattice were reported below 4 K [18]. The magnetization reversal for $T < T_{\text{comp}}$ reflects an inherent instability of the magnetic state, due to competition between the two sublattices, and points to a more complex magnetic behaviour at low temperatures [5], which deserve to be explored.

The ability to control the magnetic state of NdFeO₃ has motivated studies under oriented magnetic field [5,21]. Previous work focused on the field dependence of Fe-sublattice magnetic resonances or magnons, namely the quasi-ferromagnetic (σ-mode or qFM) and quasi-antiferromagnetic (γ-mode or qAFM) excitations, probed via THz spectroscopy [21]. These studies revealed the anisotropic character of the field-induced spin reorientation from $\Gamma_4$ to $\Gamma_2$ via $\Gamma_{24}$ above $T_2$ for fields applied along the *a*-axis (up to 5 T) [21]. Remarkably, at 140 K and 100 K, the reverse transition $\Gamma_2 \rightarrow \Gamma_4$ could not be induced by magnetic field up to 3 T applied along the *c*-axis [21]. This was attributed to the high susceptibility of the $Nd^{3+}$ moments: in the field along the *c*-axis, the Zeeman energy is insufficient to drive the system back to the $\Gamma_4$ structure, leaving $\Gamma_2$ and $\Gamma_{24}$ configurations stable [21,22]. Yet, these studies did not explore higher fields or a broader temperature range, leaving open the question of whether fields along the *c*-axis can induce magnetic phase transitions.

Motivated by these unresolved issues, we investigated the effect of high magnetic fields ($|B_{\text{ext}}|$ up to 14 T) on the magnetic states of NdFeO₃, tracking the field dependence of optically active magnon excitations using polarized THz absorption and unpolarized Raman scattering in oriented single crystals. Experiments were conducted over a wide temperature range (2-300K) encompassing the various magnetic phases of the compound. Complementary measurements were performed on the same samples, including temperature-dependent X-ray magnetic circular dichroism (XMCD) at the $Nd^{3+}$ absorption edge, magnetization and specific-heat measurements under applied fields, and torque magnetometry and isothermal magnetization at fixed temperatures, as well. This study provides detailed insights into the transformations of the magnetic structure and the interplay between $Fe^{3+}$ and $Nd^{3+}$ sublattices, culminating in the proposed magnetic phase diagram, and shed light on mechanisms that may be common to other rare-earth orthoferrites.

## 2. Field-induced magnon evolution in NdFeO$_3$ at different fixed temperatures

The Fe-spin sublattice magnons have been studied and their symmetry properties are well-established [1,4,9,23,24]. The magnetic field dependence of the magnon signatures was probed using polarized THz and unpolarized Raman spectroscopy at several fixed temperatures. The complete set of experimental spectra is presented in Section a) of the Supplemental Material, while Section b) summarizes the THz magnetic-field ($\boldsymbol{H}^\omega$) polarization-dependent selection rules and the corresponding selection rules for Raman scattering. Briefly, in THz absorption, the γ mode is active for $\boldsymbol{H}^\omega \| \boldsymbol{F}$, while the σ mode is active for $\boldsymbol{H}^\omega \perp \boldsymbol{F}$, where $\boldsymbol{F}$ is the net ferromagnetic vector arising from spin canting. The temperature dependence of the zero-field magnon modes is provided in Section c) of the Supplemental Material and serves as a reference for the field-dependent results.

The magnetic field dependence of γ mode, probed by polarized THz spectroscopy at 200 K, is shown in Figure 1. Panels (a) and (b) show the extinction coefficient spectra, κ(ω), for $\boldsymbol{B}_{\text{ext}} \| a$, for 0 and 3 T applied field strengths, respectively, with $\boldsymbol{H}^\omega \| a$ and $\boldsymbol{H}^\omega \| c$. At zero field, the γ mode is observed only for $\boldsymbol{H}^\omega \| c$. Upon increasing $\boldsymbol{B}_{\text{ext}} \| a$, its intensity progressively decreases and vanishes at 3 T (see Figures 1(b) and 1(d)), while it softens by ~0.6 cm$^{-1}$ over the 0 – 2 T interval (Figure 1(c)). Contrarily, no magnetic excitation is detected for $\boldsymbol{H}^\omega \| a$ at zero field; the mode appears at 1 T and its intensity increases with field, saturating above 3 T. These results reflect the continuous rotation of $\boldsymbol{F}$ from the $c$ toward the $a$-axis, reaching full alignment for $|\boldsymbol{B}_{\text{ext}}| > 3$ T, applied along the $a$-axis.

For $\boldsymbol{B}_{\text{ext}} \| c$, see Figures 1(e) and 1(f), the γ mode is observed for $\boldsymbol{H}^\omega \| c$ up to 7 T. The small feature observed near 18 cm$^{-1}$ for $\boldsymbol{H}^\omega \| b$ is likely a leakage resulting from a small sample misorientation. The peak observed at 10 cm$^{-1}$ is probably the σ mode, although measurement accuracy is limited at this frequency. In this configuration, the frequency of the γ mode increases by 1.6 cm$^{-1}$ as the field strength increases up to 7 T (Figure 1(g)), while its intensity remains nearly constant (Figure 1(h)). The absence of spectral-weight transfer between polarizations shows that $\boldsymbol{F}$ preserves its orientation along the $c$-axis for $\boldsymbol{B}_{\text{ext}} \| c$. Thus, at 200 K,

the $\Gamma_4$ magnetic structure, with $\boldsymbol{F}\|c$, is stable for $\boldsymbol{B}_{\text{ext}}\|c$, whereas $\boldsymbol{B}_{\text{ext}}\|a$ drives a spin reorientation process culminating in the $\Gamma_2$ phase, with $\boldsymbol{F}\|a$.

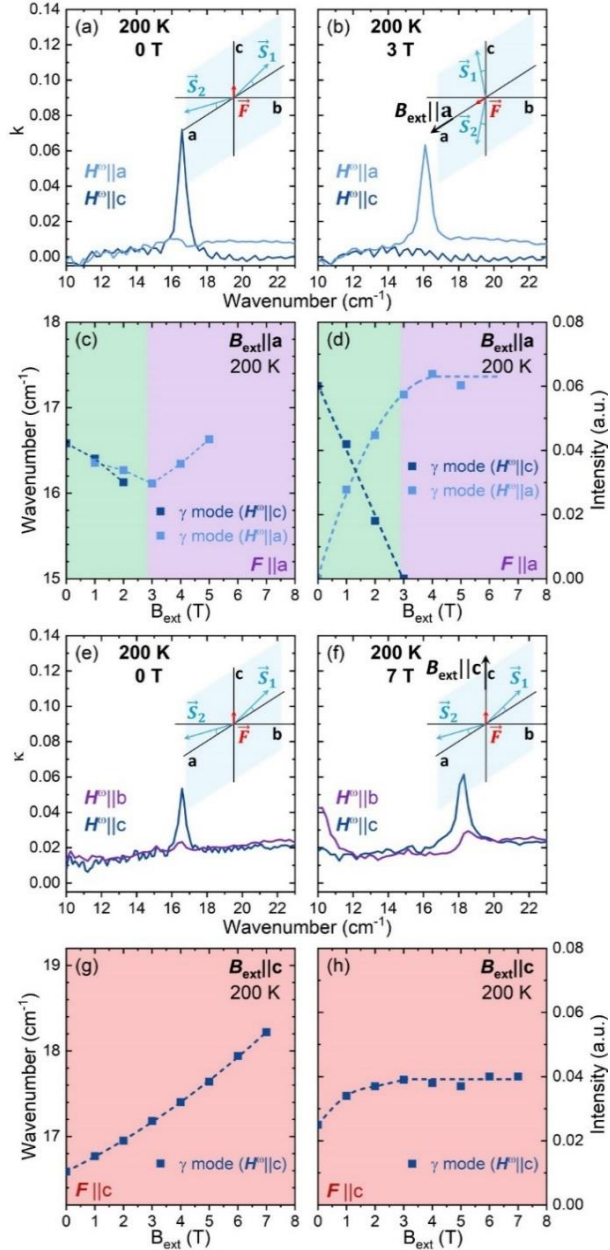

**Figure 1.** Extinction coefficient spectra of NdFeO$_3$, in different polarization configurations of the THz radiation at 200 K: (a) and (e) without applied magnetic field; under magnetic field along (b) *a*-axis and (f) *c*-axis; for $\boldsymbol{B}_{\text{ext}}\|a$: magnetic field dependence of (c) wavenumber and (d) intensity; for $\boldsymbol{B}_{\text{ext}}\|c$: magnetic field dependence of (g) wavenumber and (h) intensity. Dashed lines are a guide for the eyes. The insets of (a), (b), (c) and (f) show the orientations of the two spins of the Fe-sublattices ($\boldsymbol{S}_1$, $\boldsymbol{S}_2$) and the net weak ferromagnetic vector $\boldsymbol{F}$.

We next report the influence of an external magnetic field on magnons within the spin reorientation transition region (105–170 K), where the $\boldsymbol{F}$ vector undergoes spontaneous rotation within the *ac*-plane [5,9,25]. Figures 2(a) and 2(b) show the $\kappa(\omega)$ spectra recorded at 140 K, for the $\boldsymbol{H}^\omega\|a$ and $\boldsymbol{H}^\omega\|c$ polarization configurations, respectively, without and with an applied magnetic field along *a*-axis. Figures 2(e) and 2(f) present κ(ω) measured at 120 K for $B_{ext}\|c$, in the $H^{\omega}\|a$ and $H^{\omega}\|b$ polarization configurations. These measurements track the evolution of magnon modes as the *F* vector rotates under combined thermal and magnetic-field effects.

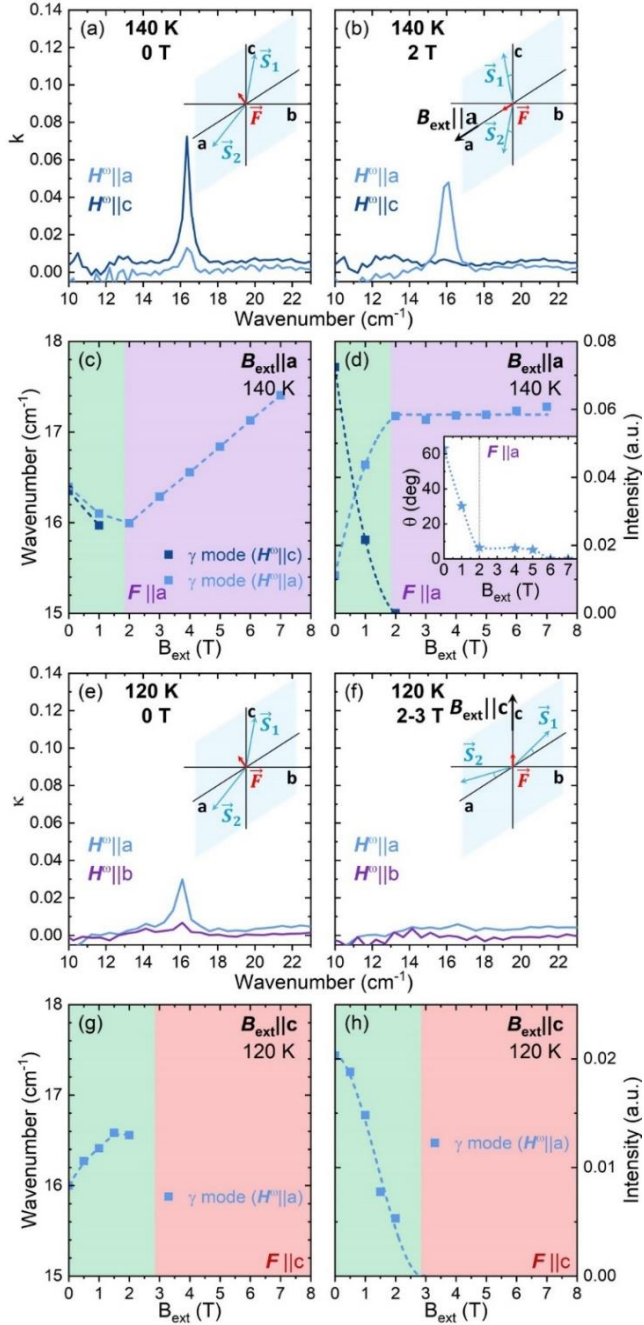

**Figure 2.** Extinction coefficient spectra, κ(ω), of NdFeO₃ recorded in different THz polarization configurations at temperatures 140 K and 120 K (within the SRT region): (a) and (e) without applied magnetic field; (b) and (f) under an external magnetic field along the *a*- and *c*-axis, respectively. Magnetic field dependence of the magnon-band (c) frequency and (d) intensity for $B_{ext}\|a$, and (g) frequency and (h) intensity for $B_{ext}\|c$. The inset of (d) shows the magnetic field dependence of the angle θ between the *F* vector and the *c*-axis, determined from Eq. (2). Dashed lines are guides to the eye.

According to the selection rules, in the spin reorientation transition temperature range at zero magnetic field, the γ mode is only observed for $H^\omega \| a$ and $H^\omega \| c$ polarizations (Figure S5, Section b) of Supplemental Material), as confirmed in Figures 2(a) and 2(e), where a single peak appears, though with different intensities. With increasing $B_{ext} \| a$, the γ mode intensity in the $H^\omega \| c$ spectrum decreases and vanishes above 2 T (Figures 2(b) and 2(d)), whereas in $H^\omega \| a$ polarization, it increases, reaching a constant value above 2 T. The corresponding wavenumber attains a minimum value at 2 T and increases by 1.6 cm$^{-1}$ up to 7 T (Figure 2(c)).

In polarized THz absorption experiments, the measured magnon intensity depends on the angle $\theta$ between the $F$ vector and the $H^\omega$ field of the THz beam, allowing $\theta(B)$ to be estimated using Eq. 2 (see Ref. [26]):

$$I_\gamma(\theta) = I_o cos^2(\theta) \qquad (2)$$

Here, $I_o$ is the maximum peak intensity, assumed to be the mean value measured for $|B_{ext}| > 2$ T, as shown in Figure 2(d) (i.e., when the $F$ vector lies along the $a$-axis), and $I_\gamma(\theta)$ is the peak intensity at angle $\theta$. At 140 K, $\theta = 64°$ at zero field and decreases with increasing $|B_{ext}| > 2$ T (inset of Figure 2(d)), evidencing a magnetically induced reorientation transition in which $F$ progressively rotates toward the $a$-axis up to the critical value 2 T.

The reverse transition, in which the $F$ vector rotates toward the $c$-axis, occurs at 120 K under $B_{ext} \| c$ (Figures 2(e) and 2(f)), completing at ≈ 3 T, as the γ mode intensity for $H^\omega \| a$ vanishes above this value (Figure 2(h)). Absence of magnon signature for $H^\omega \| b$ confirms the rotation of $F$ toward the $c$-axis (see Figure S2(b), of Supplemental Material). Contrasting with the observations reported in Ref. [21], we found supporting evidence that the phase sequence $\Gamma_{24}$ → $\Gamma_4$ is also possible at 120 K, driven by the external field applied along the $c$-axis. To confirm our spectroscopic findings, we carried out isothermal magnetization measurements, with the external field applied along either the $a$- or $c$-axes, at fixed temperatures: 200 K, 140 K, and 120 K. The results are presented in Section d) Figure S8 of Supplemental Material. Consistent with Ref. [21], the $M(B)$ curves exhibit clear antiferromagnetic signatures. For $B_{ext} \| a$ and low fields, $M_a(B)$ is a non-linear function, up to a threshold field that coincides with the critical field identified in the spectroscopic data. Above this field, $M_a(B)$ is a linear function. The low-field curvature progressively fades out upon cooling from 200 K to 120 K. By contrast, $M_c(B)$ is strictly linear above the coercive field at 200 K, but it develops a curvature at low fields on cooling to 120 K. The field-dependent evolution of magnetization is compatible with the spectroscopic data. Thus, we confirm that an external field applied in the spin reorientation temperature range, along either *a*- or *c*-axes can switch the **F** vector orientation toward the field direction above the critical field.

In the following, we address the effect of magnetic field at 110 K, 5 K above the lower temperature limit of the spin reorientation transition ($T_2$ = 105 K). Figures 3(a) and 3(d) show the magnetic field dependence of σ and γ mode frequencies and intensities measured by unpolarized Raman scattering, with the magnetic field applied along the *a*- and *c*-axes, respectively.

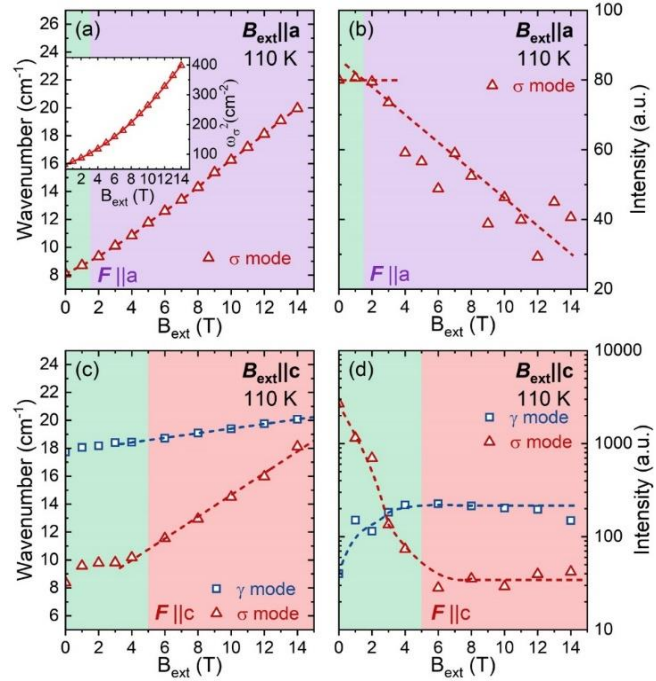

**Figure 3.** Magnetic field dependence of the σ and γ modes wavenumbers and intensities for the applied field parallel to the (a)-(b) *a*-axis and (c)-(d) *c*-axis, measured in unpolarized Raman spectra at 110 K. Inset of (a) shows the second order polynomial dependence of the squared wavenumber on the magnetic field, following the model of Ref. [27].

For $B_{ext}\|a$ (see Figure 3(a)), only the σ mode is observed, while for $B_{ext}\|c$ both σ and γ modes are detected (see Figure 3(c)). Regardless the applied field orientation, the frequencies of the observed magnons increase, though with different field dependencies. For $B_{ext}\|a$, the squared wavenumber follows a quadratic dependence on the applied field (see inset of Figure 3(a)), as predicted by the model described in Ref. [27]. For $B_{ext}\|c$, the γ mode wavenumber is a linear function of the applied field while the σ mode remains constant on increasing field up to 4 T, and then it increases linearly with further magnetic field increase. The effect of the applied field at 110 K is most pronounced for the σ mode, with a total variation of 12 cm$^{-1}$ for $B_{ext}\|a$, and 8 cm$^{-1}$, for $B_{ext}\|c$, which corresponds to a significant change of energy of 119% and 100%, respectively, associated with the Fe-sublattice spin system modifications. The magnons are observed in the entire magnetic field range, yet their intensities follow different field dependences. The intensity of the σ mode is constant up to 2 T and decreases on increasing $B_{ext}\|a$ but it is still detected at 14 T, revealing that the **F** vector is not yet fully aligned along the *a*-axis. The magnon intensities cross at 4 T and remain constant above this field. This result shows that the **F** vector reorients toward the *c*-axis (Figures 3(c) and 3(d)). At 110 K for zero-field conditions, the spontaneous reorientation is nearly complete; the **F** lies near the *a*-axis, making $B_{ext}\|a$ more effective than $B_{ext}\|c$ in modifying magnon structure.

We next focus on the effect of magnetic fields in the $\Gamma_2$ phase (50, 14, and 4 K), where the **F** vector is spontaneously oriented along the *a*-axis. Figures 4(a) and 4(b) show the results for $B_{ext}\|a$, and Figures 4(c)–4(f) for $B_{ext}\|c$.

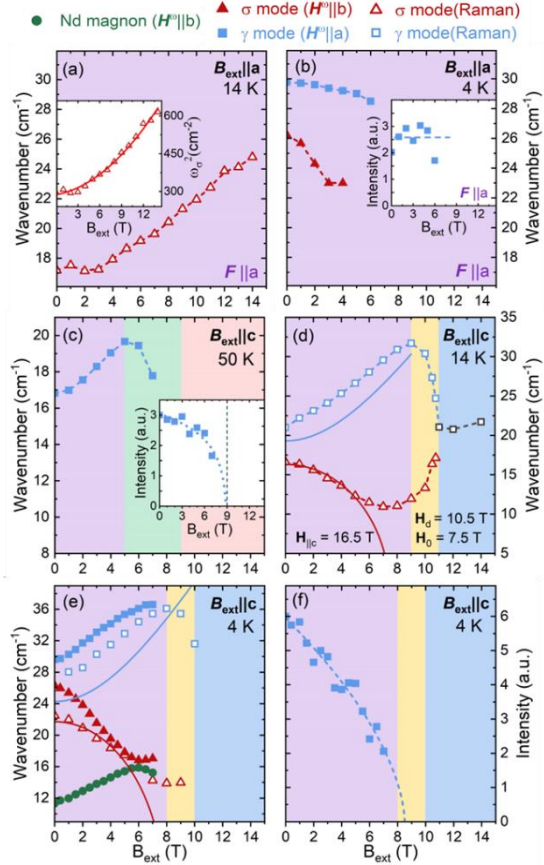

**Figure 4.** Magnetic field dependence of the σ and γ mode frequencies observed in THz (closed symbols) and unpolarized Raman (open symbols) spectroscopies at (a) 14 K and (b) 4 K for $B_{ext}\|a$. The inset of (a) shows the second-order polynomial dependence of the squared frequency on the magnetic field, following the model of Ref. [27]. Inset of (b) presented the magnetic field dependence of the γ-mode intensity. Magnetic field dependence of the σ and γ mode frequencies observed in THz and unpolarized Raman spectroscopies at (c) 50 K, (d) 14 K and (e)-(f) 4 K for $B_{ext}\|c$. The inset of (c) and panel (f) show the magnetic field dependence of the γ mode intensity at 50 and 4 K, respectively. Dashed lines are guide for the eyes. The red line represents in panels (d) and (e) were determined by the best fit of Eq. (4) to the σ mode frequency. The parameters obtained from this fit were then fixed in Eq. (3) to model the γ mode as a function of the magnetic field, shown by the blue line.

For $B_{ext}\|a$ and at 14 K, the σ mode is observed across the 0-14 T range, with its frequency squared following a quadratic dependence on the magnetic field (inset of Figure 4(a)) [27]. At 4 K, both magnons are detected up to 6 T, with the γ mode intensity remaining nearly constant throughout this field range (inset of Figure 4(b)). The results are consistent with a steady **F** orientation along the *a*-axis. Although the Fe-sublattice spin structure at zero field corresponds to the $\Gamma_2$ representation at both temperatures, the magnetic field dependence of the σ mode differs; while it hardens by 8 cm$^{-1}$ with field at 14 K, it softens by 4 cm$^{-1}$ at 4 K. The contrasting magnetic field dependencies point to distinct magnetic interactions, which result from the enhanced Nd–Fe coupling near the Nd-sublattice ordering temperature ($T_{N2}$ = 1.05 K) [19,20], modifying the Fe-magnon dynamics at low temperature, as we will discuss later.

In contrast, the magnetic field applied along the *c*-axis ($B_{ext}\|c$) induces remarkable changes in the magnetic structure, as evident in Figures 4(c) and 4(e). At 50 K, only the γ mode is observed in THz spectra, whereas at 14 K and 4 K both σ and γ modes are detected in THz and/or unpolarized Raman spectra. Across all three temperatures, the γ mode frequency initially hardens up to a certain magnetic field strength, which depends on temperature, and then softens as the field is further increased, while the σ mode exhibits the opposite behaviour at 14 K and 4 K.

At 50 K, the γ mode intensity decreases with field, extrapolating to zero near 9 T (see inset of Figure 4(c)). Since its intensity does not increase for $H^{\omega}\|b$ (see Figure S2(b), Section a) of Supplemental Material), the mode must reappear in $H^{\omega}\|c$ polarized spectra, consistent with the rotation of the **F**-vector toward the *c*-axis. At 14 K, the frequency separation between the σ and γ modes reaches a maximum near 9 T before converging. Above 11 T, only a single magnon is detected up to 14 T, as shown in Figure 4(d). At 4 K, the Fe-sublattice magnons exhibit a behaviour similar to that at 14 K: the frequency separation between the σ and γ mode is maximum at 9 T (Figure 4(e)). Although THz data were limited to 7 T, unpolarized Raman spectra (Figure S4(a), Section a) of Supplemental Material) reveal a sharp reduction in magnon intensity above 9 T, making accurate tracking beyond 10 T difficult. We will come back to the magnon signatures at 4 K later on.

For $B_{ext}\|c$ and below 110 K, the magnetic-field dependence of the magnons reveals a complex phase sequence with additional magnetic transitions beyond the spin-reorientation, whose properties depend on temperature. Coming back to Figures 4(c) and 4(d), the γ mode wavenumber exhibits similar magnetic field dependencies at 50 K and 14 K, up to the field at which the frequency reaches its maximum. This suggests that the magnetic phase sequence is the same up to this field. Above 7 T, as no data is available at 50 K, a direct comparison with the 14 K case is precluded.

The partial softening of the σ mode at 14 K, shown in Figure 4(d), points to a spin-flop (SFO) transition occurring after the spin-reorientation process. Such behaviour, characterized by the softening of the σ mode and the hardening of the γ mode, is a distinctive feature of antiferromagnets at the critical field of a SFO [27–29]. According to the two-sublattice model, the magnons frequency as a function of the magnetic field applied along the easy axis direction (c-axis in the Γ$_2$ phase) is described as follows [28]:

$$\left(\frac{f_\gamma}{g}\right)^2 \sim H_{||c}{}^2 + H_d{}^2 + 2\left[1 + \left(\frac{H_{||c}}{H_d}\right)^2\right]B_{ext}^2 \qquad (3)$$

$$\left(\frac{f_\sigma}{g}\right)^2 \sim H_{||c}{}^2 - 2\left(\frac{H_{||c}}{H_d}\right)^2 B_{ext}^2 \qquad (4)$$

Here, $f_{\gamma,\sigma}$ is the magnon frequency ($f_{\gamma,\sigma} = c\omega_{\gamma,\sigma}$, $c$ is the light speed and $\omega_{\gamma,\sigma}$ is the wavenumber), $g$ is the gyromagnetic ratio, $H_d$ is the antisymmetric Dzyaloshinskii–Moriya exchange field and $H_{||c}$ is the field at which the spin-flop transition would occur without Dzyaloshinskii–Moriya interaction [28]. Figure 4(d) shows the magnons frequencies as a function of the external magnetic field along the c-axis, for 14 K, and the corresponding fit of Eq. (4) to the experimental values of σ mode. Although the proposed model in Ref. [28] does not consider the interactions between the rare-earth and Fe-spins sublattices, which become stronger when approaching $T_{N2}$ [19,20], Eq. (4) describes well the partial softening of the σ mode up to 7.5 T, as shown in Figure 4(d), with $H_d$ = 10.5 T and $H_{||c}$ = 16.5 T. From these parameters, the critical field for the SFO transition is found to be $H_0 = \frac{H_d}{\sqrt{2}}$ = 7.5 T, that occurs when $f_\sigma = 0$. Yet, considering the obtained parameters into Eq. (3), though the increasing trend of the γ mode frequency in the 0 - 8 T range is reproduced, the predicted values are about 20% smaller than the experimental ones. Still, the value obtained for $H_d$ is of the same order of magnitude as the ones published for other orthoferrites and orthomanganites [28,30]. The σ mode analysis provides evidence for a SFO transition at 14 K when $\boldsymbol{B}_{ext}||c$ with a critical field of 7.5 T. The observation of a single magnon above 11 T in the 14 K spectrum points out for the presence of an additional magnetic phase.

## 3. Magnetic-induced phase transitions at low temperatures

To gain further insight into the phase sequence and the newly identified magnetic transition, we examine specific heat and $M_c(T)$ data measured with the magnetic field applied along the $c$-axis. Figure 5(a) show representative temperature dependences of the specific heat and DC magnetization, respectively, measured during heating under magnetic fields, in the 8 - 14 T range, applied along the $c$-axis. The measurements were performed in the 2 – 40 K temperature interval and the set of complete data is shown in Figures S11 and S12, Section e) of Supplemental Material. Figure 5(b) shows a detailed view of the specific heat in the 4 - 16 K range. Three anomalies are observed in the temperature dependence of the specific heat. The broadest, lowest-temperature anomaly exhibits a strong field dependence and it has been assigned to the two-level Schottky anomaly due to the splitting of the ground state Kramers doublet of the $^4I_{9/2}$ multiplet of $Nd^{3+}$ due to the low-symmetry crystal field [19]. The other two anomalies are sharper, but with different amplitudes. The smallest amplitude anomaly, better seem in Figure 5(b), is detected between 8 K and 15 K in all $C_p(T)$ curves, measured from 1 T to 14 T. The peak amplitude increases and shifts to higher temperatures as the magnetic field applied increases (see Figures 5(b) and Figure S11 of Supplemental Material). The largest amplitude anomaly occurs between 24 K and 28 K at magnetic fields above 8 T. This anomaly shifts to higher temperatures with increasing field. No thermal hysteresis is observed and the $C_p(T)$ curves show no additional anomalies up to 40 K, regardless of the magnetic field strength applied along $c$-axis, up to 14 T.

The smallest amplitude anomaly in $C_p(T)$ has no counterpart in the $M_c(T)$ curves, as it can be ascertained from Figures 5(a) and S11 of Supplemental Material, while the largest amplitude peak correlates with an anomaly in $M_c(T)$. As shown in Figure 5(a), this feature in the $M_c(T)$ curves evolves with magnetic field: at magnetic fields of 9 T – 10 T, the feature appears as a dip, which gradually develops into a step-like plateau for fields above 10 T. Overall, these results, together with the single magnon excitation observed above 11 T at 14 K, are consistent with a more complex magnetic phase sequence occurring below 30 K, for $|B_{ext}| < 14$ T.

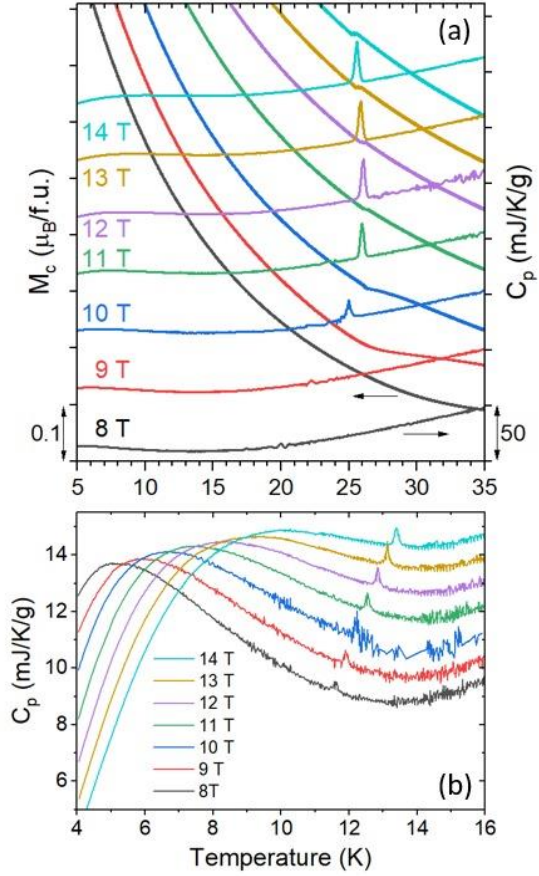

**Figure 5**. (a) Temperature dependence of the specific heat and magnetization measured for different magnetic field strengths, applied along the $c$-axis. (b) Detailed temperature dependence of the specific heat recorded at various magnetic field strengths, applied along the $c$-axis.

To characterize the phase sequence below 110 K, we have performed isothermal magnetization measurements with the magnetic field applied along the $a$- and the $c$-axis, respectively, up to 9 T, in the temperature range 2 – 100 K, and up to 14 T for selected temperatures between 4 K and 30 K. Figure 6(a) shows that the magnetization along the $a$-axis ($M_a$) increases linearly with $\boldsymbol{B}_{ext}\|a$, in the $T_{comp} <$ T $\leq$ 110 K range, confirming the absence of field-induced changes in the magnetic structure. Below $T_{comp}$ = 7.6 K, $M_a(B)$ develops a curvature reflecting a different magnetic structure, in agreement with spectroscopic observations at 4 K (see Figure 4(b)). The magnetization along the $c$-axis, $M_c(B)$, is a nonlinear function of the applied field below 110 K, as shown in Figure 6(b). The $M_c(B)$ curves exhibit a S-shape anomaly at high fields over a narrow temperature interval. To highlight this behaviour, the derivative $\frac{dM_c}{dB}$ is shown in Figure 6(c). Upon cooling from 110 to 40 K, the $\frac{dM_c}{dB}(B)$ exhibits a broad anomaly, whose amplitude increases and the maximum shifts towards higher fields. On cooling below 30 K, the anomaly narrows, shifts toward lower fields, and decreases in amplitude, disappearing at 8 K ~ $T_{comp}$.

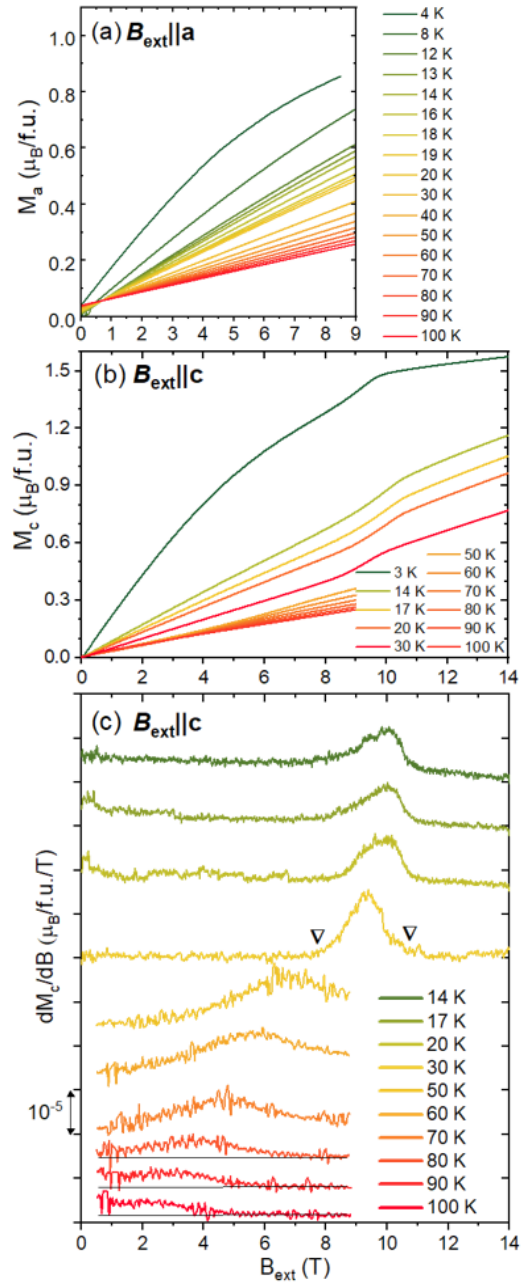

**Figure 6.** Magnetization as a function of magnetic field applied along the (a) *a*-axis and (b) *c*-axis. (c) Derivative of the magnetization along the *c*-axis. The open triangles shown in the curve measured at 30 K mark the limiting fields of the SFO transition, as representative example.

The S-shaped anomaly in $M_c(B)$ for fields applied along the easy axis (*c*-axis for temperatures below 105 K), together with the linear behaviour of $M_a(B)$, are compatible with a SFO transition [31–33], evidenced by the field dependence of magnons at 14 K for $B_{ext} \parallel c$ (see Figure 5). The SFO transition is followed by a spin-flip (SFI) transition when $M_c(B)$ returns to a linear trend, providing an estimate of the critical field, as marked by the open triangles in Figure 6(c).

The phase sequence is not the same in the 8 – 40 K temperature range. Figure 7(a) shows the magnetic field dependence of the magnetic torque at 40, 30, 20, and 14 K, as representative examples. For fields below 7 T, the torque exhibits a qualitatively similar field dependence, increasing monotonically with field. Yet, above 7 T, the torque shows markedly different field dependence at different temperatures: it rises at 40 K and 20 K but decreases at 30 K and 14 K with increasing field. The alternating behaviour, which occurs above 7 T, reflects the presence of distinct magnetic structures in the high-field regime at these temperatures. Notably, temperature dependence of the specific-heat measurements under magnetic field (see Figure 5(a)) reveal the largest amplitude anomaly between 30 and 20 K, followed by the smallest amplitude anomaly between 20 and 10 K (Figure 5(b)). The agreement between these thermodynamic signatures and the high-field torque response evidence for two distinct phase boundaries in this temperature range.

From the torque data, we obtained the transverse component of the magnetization, $M_{\text{perp}}$, which is along the $a$-axis, as a function of temperature for the different magnetic field strength applied along the $c$-axis. Figure 7(b) shows representative results, obtained for fixed magnetic field. The continuous field sweeping evolution of the magnetization is shown in Movie1 in Supplemental Material. A primary observation is the systematic reduction of $M_{\text{perp}}$ with increasing applied magnetic field. $M_{\text{perp}}(T)$ shows a strong dependence on the applied magnetic field, and two distinct field regimes can be identified in the 20 – 100 K range. For fields between 1 and 4.4 T, $M_{\text{perp}}(T)$ increases linearly as the sample is cooled from 100 K to about 60 K. the temperature rate of variation decreasing on increasing the field. Upon further cooling, the increasing rate becomes larger until $M_{\text{perp}}$ attains an almost temperature-independent plateau in the range 20–50 K. For field above 5.5 T, a different behaviour is observed. While $M_{\text{perp}}(T)$ still increases upon cooling from 100 K, a downward step appears around 60 K (see Figure 7(b)), after which $M_{\text{perp}}$ remains nearly temperature independent down to approximately 20 K. At the highest field of 8.4 T, the overall temperature trend is observed, although a shallow minimum develops around 30 K. Below 20 K, $M_{\text{perp}}$ decreases and vanishes at certain temperature and reverses direction. The temperature for which $M_{\text{perp}}$ vanishes increases from ~2 K to 6 K, on increasing field strength running from 2.4 T to 8.4 T. This behaviour contrasts with the magnetic field evolution of $T_{\text{comp}}$ under an applied field along the $a$-axis, which increases with field [5]. The magnetic field orientation dependence of the temperature for which $M_{perp}(T)$ vanishes, emphasizes the pronounced magnetic anisotropy of NdFeO$_3$. Moreover, it points to a clearly more complex magnetic state emerging below 8 K, which we ascribe to an enhanced Nd–Fe exchange interaction, as addressed in the next section.

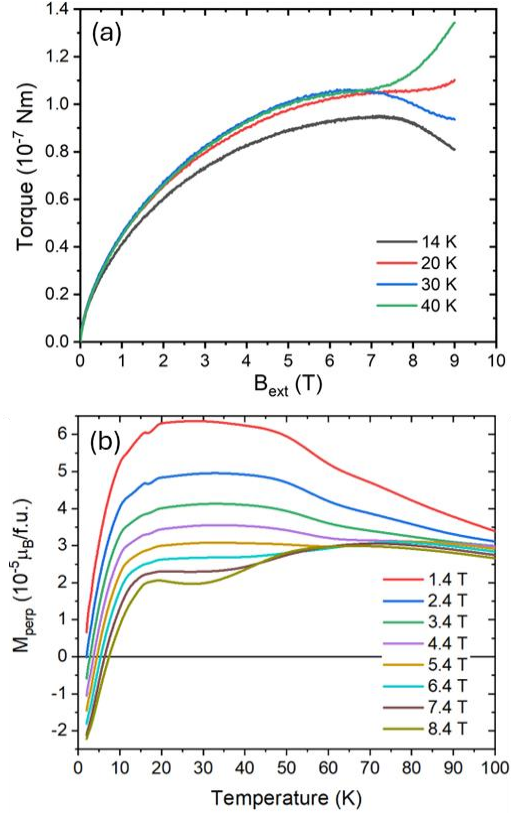

**Figure 7.** (a) Magnetic torque as a function of the applied magnetic field, measured at 40 K, 30 K, 20 K, and 14 K. (b) Temperature dependence of $M_{perp}$, calculated from the torque values, for different magnetic fields applied along the *c*-axis.

## 4. Nd-Fe magnetic crosstalk and the phase sequence below $T_{comp}$

On zero-filed cooling below $T_2$, the σ and γ modes harden [21,23], reflecting the increasing influence of the Nd-Fe magnetic interaction. As the Nd-sublattice magnetization gradually develops with decreasing temperature, as inferred from macroscopic magnetic measurements [5], its interaction with the Fe sublattice correspondingly strengthens. To confirm this behaviour, temperature dependent XMCD measurements at the Nd $M_4$ edge (~1005 eV) were performed (see Figure S13, Section f) of Supplemental Material). The squared XMCD signal, proportional to the net Nd-sublattice magnetization [34], is plotted in Figure 8 along with the temperature evolution of the Fe-sublattice magnons wavenumber shift, $\Delta\omega = \omega(T) - \bar{\omega}_{SRT}$, where $\bar{\omega}_{SRT}$ is the average wavenumber across the SRT range. The XMCD signal is negligible above 100 K. It increases monotonically on cooling as the ***F*** vector is along the *a*-axis, consistently with the strong anisotropy of the $Nd^{3+}$ 4*f* susceptibility, which is largest along the *a*-axis [13]. The parallel growth of $\Delta\omega$ and the XMCD signal underscores the magnetic coupling between the Nd- and Fe-sublattices. The coupling strength increases upon cooling below 100 K, so its effect plays an increasingly important role in the magnetic properties. This is consistent with the distinct magnetic field dependences of the Fe-sublattice magnons at 110 K and 50 K, and corroborates the different field-induced phase sequence at 110 K and at 50 K, excluding a SFO at 110 K. The rather large Nd-Fe crosstalk on approaching $T_{N2}$ = 1.05 K is the key mechanism underlying the different phase sequence observe at 14 K.

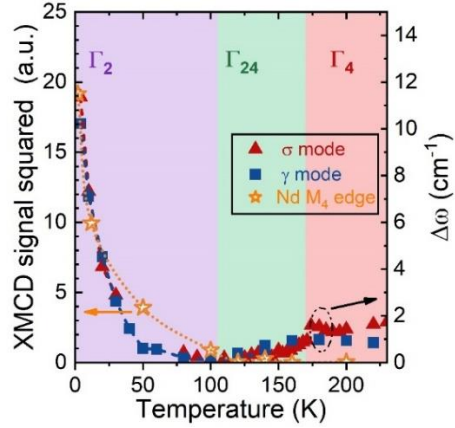

**Figure 8.** Temperature dependence of Nd $M_4$ absorption edge signal squared obtained from XMCD analysis (left axis) and the Fe-sublattice magnons wavenumber variation Δω from the average value across the SRT (right axis). The dashed lines are guide for the eyes.

Low temperature neutron diffraction studies of NdFeO₃ have revealed no signatures of intrinsic long-range Nd moment ordering above $T_{N2}$. However, the enhancement of the Nd-related diffraction intensity at (100) below 25 K has been observed and attributed to polarization of the Nd electronic magnetic moments by the Nd–Fe magnetic interaction, whereas long-range Nd-magnetic momenta ordering is only realized below $T_{N2}$ [20]. Precursors effects of the Nd-sublattice ordering have been reported. Recent studies of acoustic resonances of NdFeO₃ have shown elastic softening just below 25 K [35]. X-ray diffraction experiments also evidence lattice distortions at 40 K, ascertained by a broad but small anomaly in $b(T)$ [36]. The mechanisms of elastic softening cannot be anything other than a precursor effect of the magnetic instability associated with the Nd-sublattice ordering. The THz spectrum recorded at 4 K reveals an additional low-frequency excitation at 11 cm⁻¹ in zero field (see Figure 4(e)), which follows a magnetic field dependence closely resembling that of the γ mode of the Fe sublattice and it accounts for the partial softening of the σ mode observed at 4 K, and, eventually, at 14 K. We assign this excitation to a paramagnon arising from short-range spin correlations acting as precursor effects of the Nd-ordering transition. The emergence of Nd-sublattice magnetic excitations at 4 K and its magnetic field trend provides evidence for incipient spin ordering, accounting for the different magnetic phase sequences observed at 14 K and 4 K, and corroborate the crosstalk between the two magnetic sublattices.

## 5. Magnetostructural coupling

The progressive strengthening of the magnetic crosstalk between the two sublattices upon cooling is also reflected in the magnetic-field dependence of the Nd-oscillations probed by Raman spectroscopy (Figures S3 and S4, Section a) of Supplemental Material). The mode assignment is provided in the Supplemental Material, Section g). At 110 K, the Nd oscillations are almost magnetic field independent over the entire range explored, showing no discernible anomalies and evidencing that the applied field predominantly acts on the Fe-sublattice (Figures S14(b) and S15(c), Section g) of Supplemental Material). However, at 14 K, in-phase Nd oscillations as a function of magnetic field exhibits a slope change near 8 T, independent of field orientation (see Figure 9). The change in slope occurs at the critical field of the Fe-sublattice SFO transition at 14 K when the magnetic field is applied along the *c*-axis. A similar magnetic field dependence of the Nd-shift wavenumber is also observed for fields applied along the *a*-axis, even though no phase transition is induced in this field geometry. At 4 K, a smaller deviation appears at 4 T for $B_{ext}\|c$ (Figure S15(a)), where the torque vanishes. These findings reveal a magnetostructural coupling associated with the Nd-sublattice and underscore its distinct influence on the phase sequence below 110 K for fields along the *c*-axis.

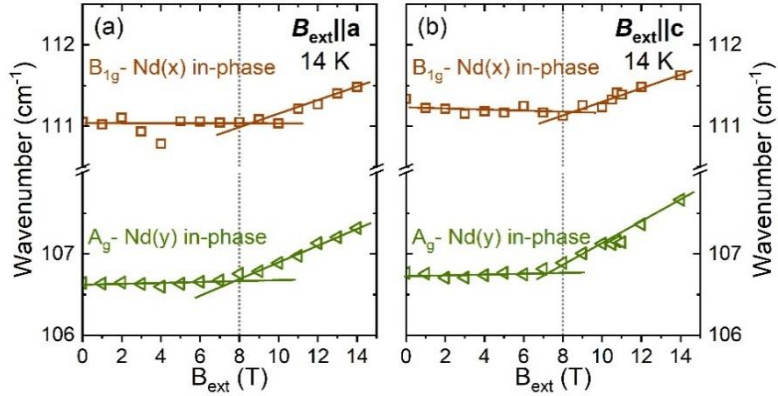

**Figure 9.** Magnetic field dependence of the Nd-oscillations wavenumbers, measured at 14 K for (a) $B_{ext}\|a$ and (b) $B_{ext}\|c$.

## 6. Magnetic phase diagram

In the following, we propose a *(B, T)* phase diagram for NdFeO$_3$ with $B_{ext}\|a$ and $B_{ext}\|c$, respectively, shown in Figures 10(a) and 10(b). Considering the phase diagram for $B_{ext}\|a$ shown in Figure 10(a), the field induced magnetic phase sequence follows as $\Gamma_4 \rightarrow \Gamma_{24} \rightarrow \Gamma_2$ from 300 K to 170 K. In the temperature range 110–170 K, the $\Gamma_{24} \rightarrow \Gamma_2$ transition occurs.

Below 100 K, no magnetic phase transition could be ascertained and the $\Gamma_2$ configuration remains stable. The critical fields obtained from $\frac{dM_a}{dB}(B)$ (see Figure S10 of Supplemental Material), resulting from the magnetic field induced SRT, allow us to draw a phase-line that is compatible with both the THz and Raman spectroscopic results. The critical field to stabilize the $\Gamma_2$ decreases as temperature decreases to 100 K.

Focusing now on the phase diagram for $B_{ext}\|c$, shown in Figure 10(b), the $\Gamma_4$ configuration is kept above 170 K. Between 100 - 170 K, the spin reorientation $\Gamma_{24} \rightarrow \Gamma_4$ occurs, with the critical field increasing as temperature decreases. In the 30 – 100 K temperature range, $\Gamma_2 \rightarrow \Gamma_{24} \rightarrow \Gamma_4$ sequence is observed. The phase sequence changes between 8 - 30 K, with the SFI transition following after the SFO. The spin structure in the SFI transition depends on temperature; at constant field, and on cooling through 10 -17 K, the small anomaly of the specific heat points for a magnetic structure rearrangement while keeping the ferromagnetic alignment. This interpretation is based on the small entropy change in the SFI-1 $\rightarrow$ SFI-2 transition. The spin rearrangement is likely due to the different role played by the Nd-sublattice magnetism on heating. The phase sequence below $T_{comp}$ = 7.6 K changes completely, yet the clear phase lines are not completely determined and are demanding additional experimental work.

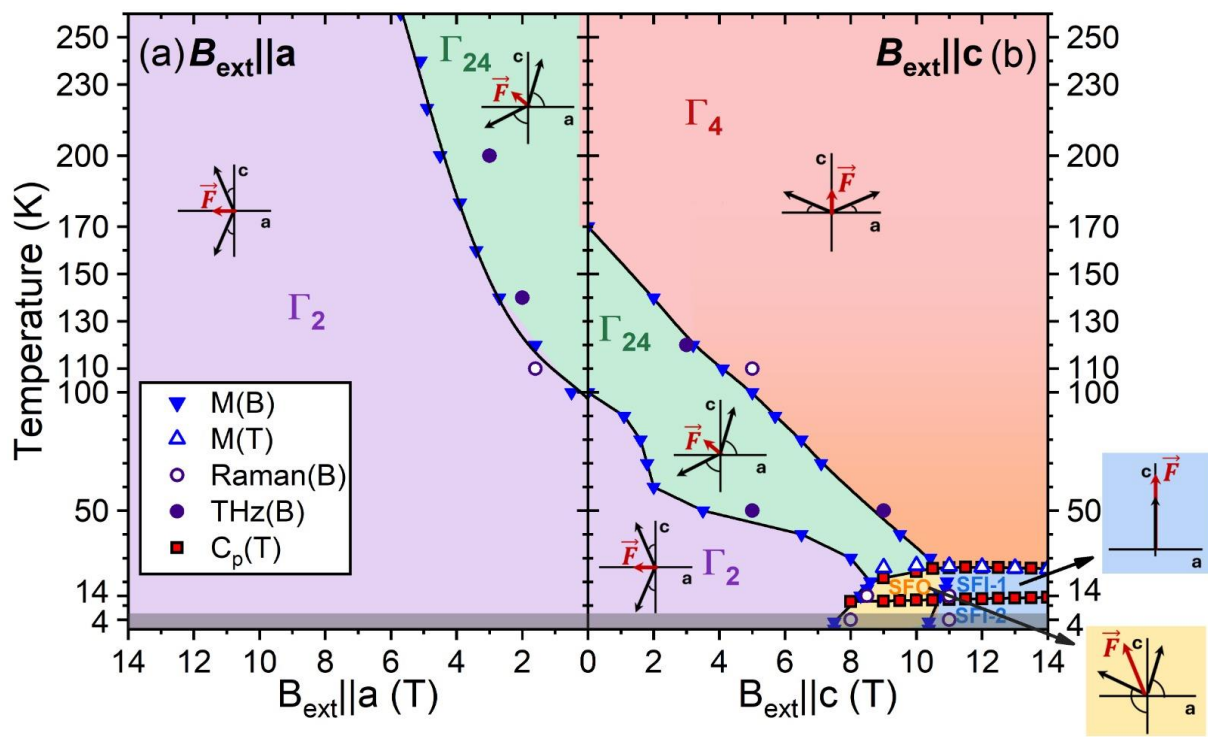

**Figure 10.** *(B,T)* phase diagram of NdFeO$_3$ for (a) $B_{ext}\|a$ and (b) $B_{ext}\|c$.

# 7. Conclusions

In this work, we presented a comprehensive temperature- and magnetic-field-dependent experimental study of the phase transitions in NdFeO$_3$. These results were obtained from polarized THz and unpolarized Raman spectroscopies, X-ray magnetic circular dichroism, specific heat, magnetic torque, and isothermal magnetization measurements.

The first main outcome concerns the progressive strengthening of the magnetic crosstalk between the Nd and Fe-sublattices upon cooling below 105 K, and its effect on the low temperature and magnetic field dependencies of Fe-sublattice spin dynamics, under an external magnetic field applied along the *a*- and *c*-axis. We identify clear signatures of magnetostructural coupling associated with the Nd-sublattice, including its influence on the complex phase sequence below 20 K, when the field is applied along *c*-axis.

The second main outcome concerns field-induced magnetic phase transition sequence and its dependence on the orientation of the applied field. Above 170 K, the phase transition sequence $\Gamma_4 \rightarrow \Gamma_{24} \rightarrow \Gamma_2$ occurs whenever the field is applied along the *a*-axis. Between 170 K and 110 K, either the $\Gamma_{24} \rightarrow \Gamma_2$ or the $\Gamma_{24} \rightarrow \Gamma_4$ transition is induced when the field is applied along the *a*- or *c*-axis, respectively. Below 100 K, the magnetic anisotropy of the Nd-sublattice and the strengthening of Nd–Fe coupling significantly impacts the induced magnetic phase sequence. For $B_{\text{ext}} \parallel c$, the $\Gamma_2$ phase transforms into the $\Gamma_{24}$ phase, followed by a spin-flop transition in the 8 – 30 K range. On further cooling and down to $T_{\text{comp}} \sim 8$ K, the spin-flop transition is followed by a spin-flip transition. Below $T_{\text{comp}} = 7.8$ K, the magnetic phase sequence is altered due to the key role of Nd-sublattice magnetism and its interplay with the Fe ssublattice. In contrast, the $\Gamma_2$ phase remains stable for $B_{\text{ext}} \parallel a$, as the magnetic field is applied along the ferromagnetic vector $F$.

The final outcome is the proposed *(B,T)* phase diagrams for NdFeO$_3$, assuming $\boldsymbol{B}_{\text{ext}} \parallel a,c$. These diagrams provide a framework for developing new models that incorporate the anisotropic character of the Nd spins, enabling a full understanding of the rich magnetic phase sequence in this compound. Furthermore, these insights can be extended to other systems with two magnetic sublattices, not limited to orthoferrites.

**Methods**

NdFeO$_3$ single crystals were grown in an optical-floating zone furnace, as described elsewhere [37]. Three sample plates were oriented along the three orthorhombic axes using Laue patterns and optically polished.

Time-domain terahertz spectroscopy experiments consisted of measuring spectra using a custom-made time-domain spectrometer [38]. For measurements without magnetic field, the spectrometer used is described in Ref. [39]. For measurements in magnetic field, broadband terahertz pulses were generated and detected using commercial fiber-coupled photoconductive switches operating with a femtosecond optical fiber laser system (TeraSmart, Menlo Systems GmbH). The oriented samples were placed in an Oxford Instruments Spectromag He-bath cryostat with mylar windows and a superconducting coil, allowing measurements in a temperature range 2 – 300 K with an applied magnetic field in the range |***B***$_{ext}$| = µ$_0$|***H***| = 0 – 7 T. The measurements used the Voigt geometry – external static magnetic field applied perpendicularly to the light propagation direction, and the Faraday geometry – external static magnetic field applied parallel to the light propagation direction.

For magnetic field-dependent unpolarized Raman measurements, a home-made experimental set-up based on free beam propagation of optical excitation and collection was used. A long working distance objective with a numerical aperture NA=0.35 and 50x ampliation was used to focus the excitation beam (λ = 515 nm) down to a spot size of 1 µm and to collect Raman scattering signal. The sample was placed on piezoelectric motors allowing for the laser focus. This set-up was inserted in a closed metallic tube filled with helium exchange gas and then placed in the middle of superconducting solenoids that produce magnetic fields up to |***B***$_{ext}$| = 14 T. The Raman scattering light was analyzed by a grating spectrometer equipped with a LN$_2$ cooled silicon Charge Coupled Device (CCD). A sum of damped harmonic oscillators was used to fit the experimental spectra [40]:

$$I(\omega, T) = [1 + n(\omega, T)]^{-1} \sum_j \frac{\omega A_{0j} \Omega_{0j}^2 \Gamma_{0j}}{\left(\Omega_{0j}^2 - \omega^2\right)^2 + \omega^2 \Gamma_{0j}^2} \qquad (1)$$

where n(ω,T) is the Bose-Einstein factor and $A_{0j}, \Omega_{0j}, \Gamma_{0j}$ are the strength, wavenumber and damping coefficient of the j-th oscillator, respectively.

X-ray Magnetic Circular Dichroism (XMCD) experiments were carried out at the DEIMOS beam line, SOLEIL Synchrotron, using the sample oriented with *c*-axis out-of-plane. To ensure optimal detection sensitivity, the absorption spectra were measured in the Total Electron Yield mode. XMCD spectra were obtained from circularly polarized absorption spectra at several fixed temperatures, with an applied magnetic field of 3 T parallel to the *c*-axis, which is also the X-ray propagation direction.

Isothermal and temperature dependent magnetization measurements were done in the temperature range from 4 to 200 K, with the applied field parallel to the *a*- or *c*-axis, from -9 T to 9 T. Detail study was performed up to 14 T in the 2 – 30 K range. The measurements were carried out using a vibrating sample magnetometer in a Quantum Design Dynacool Physical Properties Measuring System (PPMS).

The magnetic torque measurement was carried out using the torque magnetometer option in the Dynacool PPMS. A single crystalline sample was mounted on the piezoresistive cantilever attached to the torque magnetometer modulus. The sample was placed with the c-axis parallel to the applied field direction, and the b-axis oriented along the cantilever rotation axis. The angle between applied field and the *a*-axis was kept constant to 90 degrees in all measurements. The magnetic torque was measured with an accuracy of $10^{-9}$ Nm.

Heat capacity measurements were performed on a single crystal in a Quantum Design PPMS using the relaxation method. The magnetic field was applied along the *c*-axis. Data were collected between 2 and 100 K in magnetic fields up to 14 T.

**Acknowledgements**

The authors acknowledge funding from FCT and IFIMUP: NORTE-01-0145-FEDER-022096, 2022.03564.PTDC, UIDB/04968/2020, UIDP/04968/2020 projects. M.M.G. acknowledges grants from SFRH/BD/151051/2021 and R.V. from PTDC/NAN-MAT/0098/2020. M. M. and M.M acknowledge the support from VEGA 2/0004/25. Czech scientists thank the Czech Science Foundation (Project No. 24-10791S) and Project TERAFIT - CZ.02.01.01/00/22_008/0004594 co-financed by European Union and the Czech Ministry of Education, Youth and Sports. CFisUC is funded by FCT under project UIDB/04564/2025. Access to the TAIL-UC facility funded under QREN-Mais Centro project ICT_2009_02_012_1890 is gratefully acknowledged.

**Author contributions**

J.A.M. conceived the project. M.M. and M.M. grew the single crystals and M.L. cut and oriented the crystals. THz spectroscopic measurements and analysis were performed by M.M.G, C.K. and F.K. S.K. contributed to the interpretation of results. Raman spectroscopic measurements were performed by M.M.G., R.V., A.S.S. and J.A.M. under the supervision of D.J. and C.F, and respective data analysis was performed by M.M.G. and J.A.M. XMCD measurements and analysis were performed by M.M.G., R.V. and J.A.M under the supervision of F.C. Magnetization and torque measurements and data analysis were performed by M.M.G., E.M. and R.V under the supervision of J.A.P. M.M.G. and J.A.M wrote the manuscript with critical inputs from all the other authors.